\documentstyle[aps,12pt]{revtex}
\begin{document}

\title{Whispering gallery modes in open quantum billiards}
\author{R.G. Nazmitdinov$^{1,2}$, K.N.~Pichugin$^{3}$, 
I. Rotter$^1$ and 
P.~${\rm \check S}$eba$^{3,4}$}
\address{
$^1$ Max-Planck-Institut f\"ur Physik komplexer
Systeme, D-01187 Dresden, Germany\\
$^2$ Joint Institute for Nuclear Research, 141980 Dubna, Russia\\
$^3$ Institute of Physics, Czech Academy of Sciences,
Cukrovarnicka 10, Prague, Czech Republic \\
$^4$ Department of Physics, Pedagogical University, Hradec Kralove,
Czech Republic 
}
\date{\today}
\maketitle

\vspace{.2cm}

\begin{abstract}
The poles of the $S$ matrix and the wave functions of open 2D quantum 
billiards with convex boundary of different shape are calculated
by the method of complex scaling. Two leads are attached to the cavities.
The conductance of the cavities is calculated at energies with one,
two and three open channels in each lead. Bands of overlapping resonance
states appear which are localized along the convex boundary
of the cavities and contribute coherently to the conductance. These bands
correspond to the whispering gallery modes appearing in the  
classical calculations.\\


\end{abstract}

The physics of nanoscale systems has advanced rapidly over the last few years.
A consistent  description of these small systems is a 
challenging task for quantum theory 
since their properties may be influenced strongly by attaching leads to them
\cite{ben,alh,burg,berg,pepirose,seromupepi,rei1,ropepise,rei2}.
They are simulated often by means of  quantum billiards. When 
the cavity is not fully opened, the propagation of the 
mode is restricted to energies at which the overlap integral
between the wave functions of the resonance states and the channel modes
has a non-vanishing value.  In the case of well isolated 
resonances, the electron can
propagate therefore only at the energies of the resonance states
(''resonance tunneling''). 
Due to the coupling between the internal states of the cavity and the channel
mode, the states get widths. When the coupling is sufficiently strong,
the resonances start to overlap and to interact via the channel mode. 
As a consequence, some redistribution in the 
resonance states of the cavity takes place. It 
reflects the transition from the spreading of $K$ channel modes 
over the $N$ resonance states of the cavity at small opening to their free
propagation at large opening. In the last case, $N-K$ resonance states are
(practically) decoupled from the channel while only $K$ of them
are coupled strongly leading to a maximum   propagation of the $K$ waves
in the cavity. An illustration for $K=1$ is shown in \cite{rei2}.
This transition, called resonance trapping \cite{ro91},
has been studied theoretically in many different systems such as 
nuclei, atoms
and molecules (for references see \cite{ropepise}). In microwave
cavities, it is traced   as a function of the opening of the cavity 
theoretically \cite{pepirose,seromupepi,ropepise} as well as  
experimentally  \cite{perostba}. 

The role of the structure of the
cavity states themselves  in the redistribution process is almost not studied
up to now. We will show in the following that the interplay between  
the structure of the cavity states and the one-to-one alignment of
a few states with channels characterizes the situation
when the lead has a smaller extension than the cavity. Such a case
is intermediate between the two  limiting cases discussed above. 
The examples we consider
are cavities with a convex boundary which
are discussed recently in optical and other devices \cite{opt,pr}.
In these systems,  whispering gallery modes (WGM) are
known to appear  classically. We show that in the quantum mechanical
calculations, 
a certain number $N_1>K$ of resonance states  overlap, couple with comparable 
strength to the channels in both leads  while
$N_2=N-N_1$ ones are almost decoupled from the channels by resonance trapping. 
The $N_1$ states are localized near the convex boundary 
in contrast to the other ones. The localized states correspond to the WGM 
and cause an increased conduction (or reflection) of the quantum billiard.

The properties of open  quantum systems are described 
usually by means of the poles of the $S$ matrix. 
The resonance part of the $S$ matrix reads \cite{ropepise,ro91}
\begin{eqnarray}
S_{cc'} = 
i \sum_{R=1}^N \frac{\tilde\gamma_{Rc'} 
\tilde\gamma_{Rc}}{E - {\tilde E}_R + \frac{i}{2} 
{\tilde \Gamma}_R} 
\label{eq:sm} 
\end{eqnarray}
where the 
$\tilde{\cal E}_R = {\tilde E}_R - \frac{i}{2}\tilde \Gamma_R$
are the (energy dependent) eigenfunctions of  the effective Hamilton
operator 
$  {\cal H}  = {\cal H}_0  + W $ 
of the open quantum billiard, ${\cal H}_0$ is the Hamiltonian of the closed
cavity and the
\begin{eqnarray}
 W_{R'R}
 & = & \frac{1}{2 \pi} \sum_{c=1}^\Lambda {\cal P} \int
\limits_{\epsilon_{c}}^{\infty} dE' \; 
\frac{\gamma_{Rc} \; \gamma_{R'c}}{E - E'} \; - \;
 \frac{i}{2} \sum_{c=1}^\Lambda 
 \gamma_{Rc} \;  \gamma_{R'c}
\label{eq:extmi1}
\end{eqnarray}
are complex, generally,  \cite{drokplro}.
Equation (\ref{eq:sm}) is true also for  overlapping 
resonances \cite{ro91}.
The poles of the $S$ matrix correspond to the solutions of the fixed-point
equations ${\cal E}_R \equiv E_R - \frac{i}{2} \Gamma_R =  
\tilde E_R(E=E_R)- \frac{i}{2} \tilde \Gamma_R(E=E_R)$
and determine the energies $E_R$ and widths $\Gamma _R$ of the resonance
states. The  $\gamma_{Rc}$ ($\tilde\gamma_{Rc}$)
are the energy dependent coupling matrix elements  between the
eigenfunctions  $\Phi_R$  of ${\cal H}_0$ (eigenfunctions  
$\tilde\Phi_R$  of ${\cal H}$ )  and the  channel modes.
The $\tilde\gamma_{Rc}$ determine, 
together with the values $\tilde{\cal E}_R (E)$, 
the transmission   $\sum |S_{c_ic_j}^{(ij)}|^2$ 
from lead $i$ to lead $j\ne i$ and 
the reflection $\sum |S_{c_ic'_i}^{(ii)}|^2$  in the lead $i$
at the energy $E$.

In our calculations, we find the poles of the $S$ matrix and the wave functions
$\tilde\Phi_R$ of the resonance states  by using the method of exterior 
complex scaling together with the finite element method. 
For details see \cite{pepirose,seromupepi}.
The conductance is calculated from the  Schr\"odinger equation
for the closed cavity which is
modified by the boundary conditions by attaching the leads, see 
ref. \cite{ando}.

We performed  calculations for three small flat resonators of different
shape with comparable area: a semicircle  as an example for an almost 
regular cavity, a semicircle  with an internal scatterer (SIS)   
and a semi-Bunimovich billiard as examples for more chaotic 
cavities. Every resonator  is coupled to two leads, see Figs.  
\ref{fig:sis}.c -- j for the SIS.
The eigenvalue pictures are very similar to one another
in all three cases. 
There are different groups of poles (see Fig. \ref{fig:sis}.a):
one group is very near to the real axis while the other ones are deep
into the complex plane.
The states near to the real axis
are trapped by  those of the other groups.

The short-lived states start at the
opening of thresholds  lying at $E= \pi^2, \; (2\pi)^2 $ and $(3\pi)^2$. 
They form bands of 
overlapping resonance states, i.e. their widths are larger than their
distance in energy. The states of the band $A$ 
appear when the first channel opens. They are coupled strongly to 
the open channel in each of both leads (Figs. \ref{fig:sis}.c, d, g).
At $E=(2\pi)^2$, the second band ($B$) of poles 
starts whose wave functions  are related to two channels 
in each lead (Figs. \ref{fig:sis}.e, h).  Here, also another group $B1$ 
of states arises 
the widths of which increase first but then 
decrease with increasing energy (Fig. \ref{fig:sis}.f). 
The bands $A$ and $B$ continue to higher energy while a new band
($C$) starts at $E=(3\pi)^2$ where the third threshold opens. The wave 
functions of
these states are related to three channels in every lead
(Fig. \ref{fig:sis}.i). Also here, another group ($C1$) of states appears
 whose widths first increase but then decrease with
increasing energy (Fig. \ref{fig:sis}.j).

In all three cases studied by us (semi-Bunimovich and semicircle billiards),
the structure of the cavity states plays an important role in the trapping
process. The most interesting result is the 
localization of the wave functions of  the states belonging to the 
bands $A, ~B$ and $C$, called WGM in the following. 
The first WGM is pushed in direction to the border of the cavity 
at higher energies while  the second one is parallel to it. 
Even some hint to a third WGM can be seen 
when three channels are open at $E > (3\pi)^2$.
In contrast to the WGM, the long-lived states are spread over the whole cavity.
The WGM are characteristic of the structure of the states of 
the closed systems considered by us. Attaching the leads as in 
Fig. \ref{fig:sis}, they have larger coupling matrix elements
$\gamma_{Rc}$ in relation to the  channels $c$ 
than the other states. When the cavities become opened, their  widths can 
therefore increase  strongly   by trapping the other resonance states.
The states belonging to the WGM
form bands with a (nearly) square root dependence on energy.

In Fig. \ref{fig:dist}.a, we show the eigenvalue picture for the
SIS whose convex surface  is distorbed by a  cut
(Fig. \ref{fig:dist}.g). There are two groups of
short-lived states corresponding to the fact that the WGM splits, by the
distorbing  cut, into two parts with different lenghts 
of the ''ways''   for reflection in channel 1 and 2. 
The two separated parts of the WGM can be seen 
in the  wave functions of these states. 
One example is shown in Fig. \ref{fig:dist}.g.

The eigenvalue picture shown in Fig.
\ref{fig:dist}.c corresponds to the SIS with a shifted position
$s=1$ of one of the leads, see Fig. \ref{fig:dist}.h. 
Also this eigenvalue picture  shows some  band structure.
The states of $A$ split into two parts: the  localized 
part (related to the WGM) is coupled more strongly to  
the unshifted lead than to the 
shifted one while the other (not localized) part,
arising from the ''sea'' of almost bound states,  is coupled 
also to the shifted  lead. 
Both parts interact strongly at higher energy
where two channels are open.

The trajectories of the eigenvalues 
as a function of the shift $s$ of the left lead (Fig. 
\ref{fig:dist}.e) show the mechanism of resonance trapping.
The widths of the WGM decrease as a function of increasing $s$ while the
widths of some states of the ''sea'' of almost bound states
increase according to
the sum rule $\sum \tilde \Gamma_R(E) = const$. In the neighbourhood of
$s=2.5$, resonance trapping occurs  between the latter states: 
the widths of three states
become maximum at $s=2.5$ by trapping the other ones which again approach
the real axis at this value of $s$.  
The wave functions of the short-lived states  
have a clear bouncing-ball structure  (Fig. \ref{fig:dist}.i).
The poles of the WGM are
almost independent of $s$  near $s=2.5$ and their wave functions
have kept their WGM-structure  (Fig. \ref{fig:dist}.j). 
The bouncing-ball structure is less stable than the  WGM one since
the degree of 
overlapping of the poles is smaller (Fig. \ref{fig:dist}.e).

The properties of the resonance states
are reflected, at least partly, in the conductivity
of  the cavity. According to equation   (\ref{eq:sm}),
the maximum value of the matrix element $S_{cc}^{(ij)}$ is reached for 
$\tilde\gamma_{Rc}^{(i)} \approx  \tilde\gamma_{Rc}^{(j)}\; \; (j\ne i)$.  
While single short-lived states formed by resonance trapping are 
aligned each with  one  channel and
$\tilde\gamma_{Rc'} \ll  \tilde\gamma_{Rc}$, 
each of the two states $R$ and $R'$ is coupled  to both channels
with almost the same strength 
when the poles overlap (${\cal E}_{R'} \approx {\cal E}_R$) as the poles 
 of the WGM do. 
The WGM are expected therefore to cause a large conductance of the
cavity when the leads are attached symmetrically.

In Fig. \ref{fig:sis}.b, the conductivity of the SIS is
shown as a function of energy. The coherent structure 
of the conductance in the energy interval between the two lowest thresholds
can be seen clearly. 
The mean value of the conductance is about 0.9 
in the energy interval between the two lowest thresholds. 
In the next higher energy region,  
the value of the conductance is never smaller than $1$ 
meaning that the wave is reflected  into one channel, at the most.

We calculated also the conductance and reflection of the SIS, when its
shape is distorted either by a  cut in the circular boundary
of the cavity (Fig. \ref{fig:dist}.b)
or by a somewhat shifted position of one of the leads
(Fig. \ref{fig:dist}.d). 
The reflection shows, in the first case,
the same coherent structure as the conductance in the undistorted case.
The mean value is large in the energy regions considered. In the other case, 
the conductance decreases strongly with energy in the  interval between the
first and second threshold. This is caused by the fact
that  modes with higher energy move more nearly to the  convex boundary of the
cavity than those with lower energy. This result is supported by the behaviour
of the conductance at higher  energy which does not
exceed the maximum value of the conductance of a one-channel case. 
The conductance as a function of the position of one of the leads 
(Fig. \ref{fig:dist}.f) decreases strongly with approaching the 
bouncing-ball situation ($s=2.5$) in which reflection can occur only.

To shed light on the quantum mechanical results we 
consider the classical motion of a free particle inside  billiards
with the same geometry as discussed above.
The potential is assumed to be zero inside the billiard and the 
boundaries are  mirrors for the motion of the particle along
trajectories  calculated from the laws of the geometric optics. 
The dynamics of the motion 
can be reduced to a canonical mapping in  Birkhoff
coordinates $(q,p)$ \cite{Birk} which is a Poincare mapping 
at the boundary of the billiard.
The coordinate $q$ is that of the arc  length  at the boundary
of the billiard where the bounce takes
place, and $p_t={\bf {\vec p}\cdot {\vec t}/|{\vec p}|}$ is 
the tangential momentum at this point. 
Each trajectory starts at some arbitrary initial point $(x_0,y_0)$
of the attached leads with an angle $\alpha_0$ 
characterizing the direction of the motion.
The trajectories which run close to the convex boundary of our 
billiard  and accumulate upon it are defined as the
WGM of the billiard. 
These trajectories occupy the major part of the  surface of section
as can be seen from Fig. \ref{fig:cl} for the SIS.
The symmetry of the chosen geometry is
reflected in the mapping of the regions corresponding to a different number 
of bounces at the boundary. 
Additionally, we calculated the ratio of the number of trajectories 
of the  WGM to the total number of 
trajectories  passing through the billiard. This ratio is about 
$60\%$   for the semicircle and about  $70\%$ 
for the SIS, since the non-WGM trajectories are hindered by the
scatterer inside the cavity.
This result supports the idea that the WGM give the main contribution to
the conductance. Further, a billiard with  a convex boundary possesses 
a family of invariant tori which correspond to the motion close to boundary
\cite{class}. In our quantum mechanical simulations, the 
wave functions of the states of
the WGM band are localized close to the convex boundary in  the closed 
systems and remain localized when the systems are opened 
to a few channels by attaching leads to them. 
Thus, the classical results are in qualitative
agreement with the quantum-mechanical ones.

\vspace{.2cm}
{\it Summarizing the results}, we state the following. 
In open quantum cavities with convex boundary, 
the structure of the wave functions of the cavity states plays an important
role in the trapping process. Some  resonance
states  receive large widths by trapping the remaining  ones, 
form bands of overlapping resonance states and are localized near 
the convex boundary of the cavity.
These bands correspond to the WGM known from classical calculations
for different  systems with convex boundary.
When two leads are attached symmetrically to the cavity,
the WGM are coupled to both leads with comparable strength.
WGM-like structures exist in relation 
to every open  channel in the two waveguides. 
The WGM are responsible for a high 
conductivity which decreases dramatically when the symmetry of the system 
is distorted. This fact can be used surely for the design of quantum
cavities in practical applications.
\\

\vspace{.8cm}

{\it Acknowledgment:} K.P. and P.S. gratefully 
acknowledge the hospitality of the
MPI PKS   during their stays in Dresden.

\vspace{.8cm}
\noindent
e-mail: 
~~rashid@thsun1.jinr.ru; ~~pichugin@fzu.cz; \\
rotter@mpipks-dresden.mpg.de;
~~petr.seba@uhk.cz

\vspace*{2cm}

\begin{figure}
\caption{
The poles of the $S$ matrix (a) and conductance 
$G=\sum |S_{c_ic_j}^{(ij)}|^2$ (b)
for the  SIS. The poles of the $S$ matrix (denoted by stars)
are  connected by lines for guiding the eyes. The full lines in (b)
show the mean value of the conductivity 
between every two thresholds. Some wave functions ($|\tilde\Phi_R|^2$)  
of the SIS: 
(c) from band $A$ of the first energy interval
$\pi^2<E<(2\pi)^2$, (d), (e), (f) from bands $A$, $B$, and $B1$ 
of the second energy
interval $(2\pi)^2<E<(3\pi)^2$, (g), (h), (i) and (j) 
from bands $A$, $B$, $C$ and $C1$
of the third energy interval $(3\pi)^2<E<(4\pi)^2$.
The energies $E_R$ and widths $\Gamma_R$
are in units  $ \hbar^2 / (2 m d^2)$ where $d=1$ is
the width of the attached waveguide.   
}
\label{fig:sis}
\end{figure}

\begin{figure}
\caption{
The poles of the $S$ matrix  (a, c), 
reflection $R=\sum |S_{c_ic'_i}^{(ii)}|^2$ (b) and 
conductance $G=\sum |S_{c_ic_j}^{(ij)}|^2$ (d)
for the  SIS distorted by a cut in the 
circular boundary of the cavity (a,b) and by a somewhat 
shifted position of one of the leads (c,d). 
One wave function ($|\tilde\Phi_R|^2$) for each case is shown in (g) and (h).
The trajectories of the poles of the $S$ matrix (e) and 
the mean value $\overline G$ of the 
conductance  (f) as a function of  the position $s$ of the left lead.
In (e), the poles at the positions $s=0, 2.5, 3$ are marked with $\Diamond$,
$\bigtriangleup$ and $\bigtriangledown$.
Two wave functions (i, j)   at $s=2.5$.
For further explanations see Fig.\ref{fig:sis}.
}
\label{fig:dist}
\end{figure}

\begin{figure}
\caption{
Classical Poincare section for the SIS
in the Birkhoff variables: tangential momentum $p_t$ vs.  arc length $q$.
The trajectories of the WGM are distinguished
according to the number $m$ of bounces at the boundary
for $m \le 5$.
}
\label{fig:cl}
\end{figure}

\end{document}